# A Complete Transport Validated Model on a Zeolite Membrane for Carbon Dioxide Permeance and Capture

Evangelos I. Gkanas[1,2*], Theodore A. Steriotis[3], Athanasios K. Stubos[2], Sofoklis S. Makridis[1,2]

[1] Institute for Renewable Energy and Environmental Technologies, University of Bolton, Deane Road, Bolton, UBL3 5AB, United Kingdom

[2] Institute of Nuclear Technology and Radiation Protection, NCSR "Demokritos", Aghia Paraskevi, Athens, 15310, Greece

[3] Environmental Technology Laboratory, NCSR 'Demokritos', Agia Paraskevi, Athens 15310, Greece

*E-mail address*: gkanas@ipta.demokritos.gr (E.I Gkanas).

**Abstract**

The $CO_2$ emissions from major industries can cause serious global environment problems and their mitigation is urgently needed. The use of zeolite membranes is a very efficient way in order to capture $CO_2$ from some flue gases. Zeolite membranes are porous crystalline materials with pores of a consistent size and these pores are generally of molecular size (0.3 to 1.3) nm, and enable high selectivity and reduced energy requirements in industrial separation applications. Further, zeolites are thermally stable and have known surface properties. Separation in zeolites is mainly based on dissimilarity of diffusivities and favored absorption between the components.

The current work is aimed at developing a simulation model for the $CO_2$ transport through a zeolite membrane and estimate the diffusion phenomenon through a very thin membrane of 150 nm in a Wicke-Kallenbach cell. This apparatus have been modeled with Comsol Multiphysics software. The gas chosen to insert in the retentate gas chamber is $CO_2$ and the inert gas is Argon. The Maxwell-Stefan surface equations used in order to calculate the velocity gradients inside the zeolite membrane and in order to solve the velocity profile within the permeate and retentate gas chamber, the incompressible Navier-Stokes equations were solved. Finally, the mass balance equation for both gases solved with the mass balance differential equations. Validation of our model has been obtained at low and high temperatures suggesting that higher the temperature the better the results.

1. Introduction

There is a growing consensus among the scientific community that the rising atmospheric levels of $CO_2$ as a result of man-made activities, such as emissions from major industries (power generation, steel and cement industries) [1] are that the origin of the warming effect of the climate [2]. Membrane processes appear to be an attractive option to carry out gas separations in terms of their lower environmental impact and energy cost, compared to more conventional separation technologies. Furthermore, the modular nature of membranes constitutes a positive input [3].

A wide variety of micro- and meso-porous materials are of potential use in separation applications such as $CO_2$ capture [4-6]. Examples of microporous materials include zeolites (crystalline aluminosilicates) among other materials such as metal-organic frameworks, covalent organic frameworks and zeolitic imidazolate frameworks. Zeolites are inorganic crystalline structures with uniform pores of molecular dimensions. Different pore sizes and composition of zeolites have been used to prepare membranes, and zeolite membranes with different shape have been investigated to separate $CO_2$ [7-9]. These materials have unique properties such as a singular pore diameter, well-defined surface properties and high thermal stability making them invaluable in many technical applications [10, 11]. Zeolites can be applied as powders, pellets and as thin films grown on inert support membranes with a larger pore size [12]. In between, numerous zeolite membrane preparations are reported and substantial progress can be stated, examples are the preparations of zeolite membranes of types LTA [13], FAU [14], CHA [15], DDR [16] and mixed tetrahedral-octahedral oxides [17, 18]. Many researchers have successfully prepared zeolite membranes. Recently, Tsapatsis et al. [19] have prepared a b-oriented MFI silicalite-1 membrane, and they further showed the performance of h0h and c-oriented silicalite-1 layers at different temperatures [20, 21].

For the applications described, migration or diffusion of sorption molecules through the pores and cages within the crystal structure of the zeolite plays a dominant role. Configurationally diffusion is the term coined to describe diffusion in zeolites and it is characterized by very small diffusivities ($10^{-8}$ to $10^{-14}$) $m^2/s$ [22] with a strong dependence on the size and shape of the quest molecules [23, 24] and high activation energy [25]. Further, it is characterized by very strong concentration dependence [26]. The measurement of the diffusivity in zeolites is made by both macroscopic and microscopic methods.

Diffusion of components, especially gases, through porous materials can be experimentally studied with the use of Wicke-Kallenbach (W-K) cells [27, 28]. This kind of cells consists of two flow-through components separated by a membrane of porous material through which components can penetrate. A steady gas stream with certain composition, flows through one compartment, while another stream of usually inert gas flows through the other compartment as a sweep gas. Mass transport parameters of component in a porous material are frequently determined from a transport model developed to explain experimental observations with a W-K cell.

Different models have been proposed to describe various transport mechanisms in various inorganic membrane materials [29, 30].

Theoretical approaches for modeling the diffusion in zeolites and/or other microporous structures fall into two different categories. A kinetic approach and an approach based on irreversible thermodynamics. The former is based on random walk models and/or transition state theory appropriately modified to account for several additional phenomena such as multilayer adsorption, surface heterogeneity and energy barriers [31-33]. The irreversible thermodynamic approach considers the chemical potential gradient as the driving force for diffusion [34]. Multicomponent interactions occur through competitive equilibrium and/or diffusional sorbate-sorbate interactions. In this case, the driving force exerted on any particular species is balanced by the friction this species experiences with the other species present in the mixture and can be accurately described by the generalized Maxwell-Stefan model for diffusion [35].

The current study is aimed at the development of a complete simulation model for single component mass transport through a zeolite membrane in a Wicke-Kallenbach (W-K) permeation cell. A three-dimensional permeation cell was built and used to simulate velocity, flux and concentration profiles of the compartment. The generalized Maxwell-Stefan formulation has been applied successfully to describe transient and stationary uptake of carbon dioxide in zeolites. The fractional occupancies of the species ($CO_2$) are described by the Langmuir-type adsorption model.

## 2.  Maxwell-Stefan theory for diffusion through zeolite membrane

The generalized Maxwell-Stefan (GMS) equations have successfully been applied to many systems to describe diffusive transport phenomena in multicomponent mixtures and single component species [36]. These models mainly based on the principle that in order to cause relative motion between individual species in a mixture, a driving force has to be exerted on each of the individual species. This driving force is balanced by the friction this species experiences with the other species in the mixture and the friction between the species and the surface of the membrane. Krishna et al. [36] described the diffusion through the membrane of adsorbed molecules starting from the equation for an *n*-component mixture:

$$-\nabla \mu_i = RT \sum_{\substack{j=1 \\ j \neq i}}^{n} \theta_j \frac{u_i - u_j}{D_{ij}^s} + RT \frac{u_i}{D_i^s}, i = 1, 2, \ldots n \quad (1)$$

where $-\nabla \mu_i$ is the force acting on species i tending to move along the surface with velocity $u_i$. The first term on the right-hand side describes the friction exerted by adsorbate *j* on the surface motion of species *i*, each moving with velocities $u_j$ and $u_i$ with respect to the surface, respectively. The second term reflects the friction between the species *i* and the surface. $D_{ij}^s$ and $D_i^s$ represent the corresponding Maxwell-Stefan diffusivities and $\theta_j$ is the fractional surface occupancy. The GMS formulation, Eq. (1) has been applied successfully to describe transient uptake in zeolites and carbon molecular sieves, and in zeolitic membrane permeation. Generally, a multicomponent Langmuir-type adsorption model is used to describe the fractional occupancies. For thermodynamic consistency, however the saturation loading for all species must be equal in the multicomponent Langmuir model hence the fractional occupancies can be defined as:

$$\theta_i = \frac{q_i}{q^{sat}} \quad (2)$$

This implies that for different molecules different amounts are needed to obtain similar levels of fractional accupancies. To develop expressions for the diffusional fluxes through the zeolite membrane from Eq. (1), the driving force, the gradient of the thermodynamic potential of a component is related to gradient in loading through the partial pressure and the mixture adsorption isotherm. The fractional occupancies are converted into fluxes using Eq. (3).

$$N_i = \rho q^{sat} \theta_i u_i = \rho q_i u_i \quad (3)$$

In this case, the ideal adsorbed solution (IAS) theory as proposed by Myers and Prausnitz (1965) will be used, which is thermodynamically consistent and can be applied using single component isotherms. Dropping the superscripts for the surface diffusivities in the GMS expression for convenience, multiplication of both sides by $\theta_i/RT$ and application of Eq. (2), Eq. (1) can be re-written as:

$$-\frac{\theta_i}{RT}\nabla\mu_i = \sum_{\substack{j=1 \\ j\neq i}}^{n}\theta_i\theta_j\frac{u_i-u_j}{D_{ij}} + \frac{\theta_i u_i}{D_i} = \sum_{\substack{j=1 \\ j\neq i}}^{n}q_iq_j\frac{u_i-u_j}{q_i^{sat}q_j^{sat}D_{ij}} + \frac{q_i u_i}{q_i^{sat}D_i} \tag{4}$$

Using the definition of fluxes, Eq. (4) can be written as:

$$-\rho\frac{\theta_i}{RT}\nabla\mu_i = \sum_{\substack{j=1 \\ j\neq i}}^{n}\frac{q_j N_i - q_i N_j}{q_i^{sat}q_j^{sat}D_{ij}} + \frac{N_i}{q_i^{sat}D_i}, i=1,2....n \tag{5}$$

The gradient of the thermodynamic potential can be expressed by terms of thermodynamic factors [37]:

$$-\rho\frac{\theta_i}{RT}\nabla\mu_i = \sum_{j=1}^{n}\Gamma_{ij}\nabla\theta_i, \Gamma_{ij} \equiv \frac{\theta_i}{p_i}\frac{\partial p_i}{\partial \theta_j} \tag{6}$$

where Eq. (6) represents a thermodynamic factor which can be determined by the adsorption isotherm, chosen to relate the surface coverage $\theta_i$ to the partial pressure $p_i$. In the current work the adsorption is described by the extended Langmuir model, Eq. (2).

Eq. (5) and Eq. (6) can be cast in a matrix-vector relation:

$$-\rho[\Gamma](\nabla\theta) = [B][q^{sat}]^{-1}(N) \tag{7}$$

where $[q^{sat}]^{-1}$ is a diagonal matrix of saturation loadings and the elements of $[B]$ are given by the following equations:

$$B_{ii} = \frac{1}{D_i} + \sum_{\substack{j=1 \\ i\neq j}}^{n}\frac{\theta_i}{D_{ij}} \tag{8}$$

and

$$B_{ij} = -\frac{\theta_j}{D_{ij}} \tag{9}$$

The solution of Eq. (7) for the diffusion fluxes is the:

$$(N) = -\rho[q^{sat}][B]^{-1}[\Gamma](\nabla\theta) \tag{10}$$

For pure species, which in the current study is $CO_2$ Eq. (10) can be written as follows:

$$N_t = -\frac{\rho q_i^{sat} D_i \nabla \theta_i}{1 - \theta_i} \qquad (11)$$

### 3. Model Description

*3.1 Geometry of the model*

A three-dimensional model has been developed to resolve the flow pattern and $CO_2$ concentration in the W-K compartment. The development of a three-dimensional model is very important in this case because in the introduction of the Partial Differential Equations which describes the flux of $CO_2$ within the COMSOL Multiphysics is performed with matrices as already discussed with the Eq. (10). With a three-dimensional model a 3x3 matrix describing the $CO_2$ flux can be used in order to be able to calculate the flux in all the possible directions within the membrane. The zeolite membrane is very thin (approximately 150 nm thick). The cell is cylindrical in shape with a diameter of 19 mm and consists of a retentate gas chamber where the gas $CO_2$ enters the chamber, and a permeate gas chamber where an inert gas (in this case Argon) enters the system. The two chambers are separated by a cylindrical, solid zeolite membrane, usually held within a support system. In order to simplify the current problem some assumptions were taken into account. These assumptions are:

1) The transport of the absorbing components through the zeolite membrane occurs due to surface diffusion, described by the generalized M-S model.

2) Additional contributions such as gas translation are negligible

3) The pressure drop along each compartment is assumed to be negligible

4) The deformation of the zeolite membrane under high pressure is negligible

5) No support layer is taken into account [38]

6) The sweep gas does not experience counter diffusion through the zeolite membrane

Both gas chambers are 0.3mm thick. This counter-current system feeds a concentrated gas-flow into the retentate gas chamber and the chemical species reach the zeolite membrane. A portion of the species diffuses through the zeolite and is removed from the permeate chamber by feeding an inert sweep. The geometry of the model is illustrated in Figure 1.

*Figure 1. Geometry of the Wicke-Kallenbach cell*

The gas flowing in the compartment was modelled using the incompressible Navier-Stokes equations, assuming that the gas flowing in the compartment is in the laminar flow regime. The general equation that defines the incompressible flow is given by:

$$d\frac{\partial u}{\partial t} - n\nabla^2 u + \rho(u\nabla)u + \nabla p = F \qquad (12)$$

Furthermore, the mass transport in the compartment is due to both convection and diffusion. The momentum balance equation in the retentate compartment is given by:

$$\frac{\partial c}{\partial t} + \nabla(-D\nabla c) = R - u\nabla c \qquad (13)$$

### 3.2 Boundary Conditions

The appropriate boundary conditions for the solution of the problem are described by the following set of equations.

#### 3.2.1 Maxwell-Stefan diffusion through the membrane

The conditions refered to the zeolite membrane domain boundaries, found between the permeate and retentate gas chambers are: The Neumann boundary condition which refers to the edges where no flux occurs and the Dirichlet boundaries, which are used at the interface between the zeolite membrane surface and the respective permeate and retentate gas chambers. The Langmuir isotherm is used in the current study in order to calculate the surface coverage of sites at the interface between the gas chambers and the solid zeolite membrane. Thus this relationship is defined in the Dirichlet boundary conditions.

#### 3.2.2 Mass Balance

The boundary conditions between the gas and the walls of the chamber have been set as:

$$nN = 0, N = -D\nabla c + c\vec{u} \text{ (Insulation/Symmetry condition)} \quad (14)$$

The boundary condition between the gas and the membrane interface is given by:

$$-D \cdot \nabla c + c\vec{u} = N_0, \quad (15)$$

where $N_0$ is the inward flux (mol/m$^2$s)

## 4. Results and Discussion

### *4.1 Model Verification*

In order to validate the model which is proposed in the current study, a comparison between the results extracted from the simulation runs based on the model and experimental results obtained from already published data will be performed. As other groups have published their simulation results [37-38] without taking into account the effect of the support resistance on permeation and ultimately analyze the permeation behavior using adsorption and occupancy dependent diffusion within the membrane, in the current study the membrane is treated without support. Himeno et al. [39], performed adsorption isotherms of carbon dioxide at temperatures of 273, 298, 323, 348 K, at a pressure range between 0 and 3500 kPa. In order to compare the simulation results with the experimental data, simulation runs were performed at the same temperatures (273, 298, 323, 348 K). For each temperature, results of $CO_2$ concentration (mol/kg) obtained for individual pressures with a pressure step of 200 kPa, in order to cover the range of 0-3500 kPa. Finally, the results collected from these simulation runs are compared with the experimental results by Himeno et al. in Figure 2a. A good agreement between the results is obtained. Further, for lower temperatures, the same process was performed in order to compare the simulation results with the experimental data obtained by van der Bergh et al. [40]. Van der Bergh et al. obtained their results at temperatures of 298, 273, 252 and 195 K in a pressure range between 0 and 100 kPa. Again, for each temperature, simulation runs were performed for pressure steps of 10 kPa to cover the pressure range 0-100 kPa.

The comparison of the results is presented in Figure 2b. According to these data is obvious that for the temperatures of 298 and 273 the results of the simulation runs are in good agreement with the experimental data, while for the lower temperatures there is a small deviation between the results, but the shape of the isotherms is almost the same. This could be due to the fact that the support of the membrane might play a major role of thermal insulator at lower temperatures.

*Figure 2. Comparison of experimental results extracted from recently published adsorption data by Himeno et al. and the simulation results extracted at same temperatures with a pressure step of 200 kPa (2a), and experimental adsorption data recently published by van der Bergh et al. and simulation results extracted at same temperatures with a pressure step of 10 kPa (2b).*

### *4.2 $CO_2$ permeation through the zeolite membrane*

For single-component diffusion, transport flux of the component through the zeolite membrane is described by Eq. (11). The term $D_i$ is referred as M-S surface diffusivity. In the current study, three different M-S diffusivity term scenarios will be considered and compared. In the weak confinement scenario the M-S diffusivity is to be independent of the fractional occupancy and equal to the initial zero-loading diffusivity:

$$D_i = D_i(0) \qquad (16)$$

In the strong confinement scenario, the M-S diffusivity presents linear dependence of the occupancy:

$$D_i = D_i(0)(1-\theta_i) \qquad (17)$$

Reed and Ehrlich [41] proposed a general occupancy dependence on the M-S diffusivity term:

$$D(\theta) = D(0)\frac{(1+\varepsilon)^{y-1}}{(1+\frac{\varepsilon}{f})^y} \qquad (18)$$

where y is the coordination number (the maximum number of neighbors in the lattice cavity of the membrane) and the other parameters are given by:

$$f = \exp(\frac{\delta E}{RT}) \quad (19) \qquad \varepsilon = \frac{(\gamma - 1 + 2\theta)f}{2(1-\theta)} \quad (20) \qquad \gamma = \sqrt{1 - 4\theta(1-\theta)(1-\frac{1}{f})} \quad (21)$$

Figure 3 shows the transient permeation of pure $CO_2$ through the zeolite membrane for three different feed pressures (10-100 and 1000 kPa) and compared for the three different M-S diffusivity scenarios described above. All the measurements were performed at temperature 298 K, and the parameters used are presented in Table 1. The fluxes at the feed and the permeate sides are decreasing and increasing respectively until they reach steady state. As the feed pressure increases from 10 to 1000 kPa, there are larger differences among the transient fluxes for the three M-S diffusivity scenarios. However, even at the higher pressure of 1000 kPa the difference between the strong confinement scenario and the quasi-chemical approach is small and this is a consequence of the close diffusivities of $CO_2$ obtained from these two estimation methods. The same approach was performed by Lee [37] who also pointed that as the feed pressure is increasing, larger differences in the fluxes between the different scenarios are taking place, where the strong confinement scenario with the quasi-chemical approach, have almost the same behavior, even at high pressures. The main difference (which seems to be minor) between the strong confinement scenario and the quasi-chemical approach lies on the fact that the strong confinement scenario depends directly to the fractional occupancy which related with the number of the remaining free spaces for $CO_2$ capture within the membrane with time, while the quasi-chemical approach depends on more parameters except the occupancy, such as the number of adjacent atoms near the available cavity which might affect the behavior of the membrane. The weak confinement scenario on the other hand does not takes into account the fractional occupancy indicating that the Maxwell-Stefan diffusivity term remains constant with time. This is probably the main reason that the results extracted from this scenario, doesn't match to the results extracted from the other two scenarios at high pressures. The conclusion of the above results is that in high feed pressures the weak confinement scenario might bring wrong estimated M-S diffusivities while the strong and quasi-chemical approach seems to have better potential in describing the diffusion at higher pressures. At low pressures it seems that all three scenarios can describe with detail the $CO_2$ diffusion through the zeolite membrane.

*Table 1. Parameters used for the comparison of the three scenarios [37]*

*Figure 3. Transient response to an increase in feed pressure in a $CO_2$ permeation system. Feed and Permeate fluxes for the three different scenarios about the M-S diffusivities for feed pressures a) 10 kPa b) 100 kPa and c) 1000 kPa.*

### 4.3 Effect of Temperature and Pressure on the $CO_2$ permeation through the membrane

The temperature effect on the diffusion through a zeolite membrane is very important and the temperature dependence of the permeance of a large number of gases has been well-studied. Figure 4, presents the simulation results for the flux of $CO_2$ through the membrane at a temperature of 303K and a pressure range from 0 to 600 kPa at temperature 303 K. It is obvious from the results that the flux of $CO_2$ through the zeolite membrane presents a non-linear dependence on the pressure.

*Figure 4. $CO_2$ flux as function of the feed pressure at constant temperature 303 K.*

Figure 5, shows the adsorption isotherm of carbon dioxide in the zeolite membrane at 298, 273, 252 and 195 K for a pressure range from 0 to 100 kPa. It is clearly seen from Figure 5 that the isotherm changes from a non-linear to an almost linear shape if the temperature is increased. The reason of this behavior will be discussed later.

*Figure 5. Adsorption isotherms of carbon dioxide at 195, 252, 273 and 298 K.*

When the mass transport through the zeolite membrane is described the flux and the permeance are typically used. The permeance is calculated by from a mass balance at steady state by using the pressure difference between the retentate and the permeate side. The permeance can be defined as:

$$\Pi = \frac{N}{\Delta p} \qquad (22)$$

where $\Delta p$ is the pressure difference of $CO_2$ over the membrane. Sometimes permeance is better quantity to describe the mass transport, because it takes into amount the pressure difference due to some pressure variation problems that might occur due to the sweep gas diffusion mechanism within the membrane. The effect of the isotherm shape on the behavior of the permeation is illustrated in Figure 6. Figure 6a presents the $CO_2$ flux as a function of pressure and Figure 6b the permeance as a function of pressure. The parameters used for these temperatures are presented in Table 2. The simulation results for both the permeance and flux are in great agreement with the experimental results presented by van der Broeke et al. [8], and the comparison can be seen in Figure 7.

*Figure 6. Effect of pressure and temperature on the permeation of carbon dioxide through the zeolite membrane.*

*Table 2. Langmuir parameters for the temperatures 303, 363 and 423 K.*

*Figure 7. Comparison of the results extracted from simulation runs and the experimental results extracted from van der Broeke et al. [8] for the effect of pressure and temperature on the permeance.*

For the lower temperature 303K both the flux and the permeance present a non-linear behavior on the feed pressure, but this behavior is changing for higher temperatures 363 and 423 K where the permeance for the temperature 423 K is almost constant with pressure. These results are in very good agreement with the equilibrium isotherms that presented in Figure 5. For an almost linear isotherm the flux show linear behavior and the permeance seems to be independent of the pressure. This can be explained as follows. The flux through a zeolite membrane is a function of the diffusivity of the component and the amount of the component adsorbed within the zeolite. Diffusion in zeolites is an activated process and the diffusivity increases with the temperature, while the adsorbed amount decreases with the temperature. When decreasing the temperature in zeolite reaches saturation and the decrease of diffusivity begins to dominate due to the asymptotic approach of adsorption saturation.

### *4.4 Transient analysis of $CO_2$ permeation through the zeolite membrane*

For the transient analysis of the $CO_2$ permeation through the zeolite membrane it is expected that the $CO_2$ flux through the membrane initially will be increased and after some time the equilibrium situation will be reached. Further, as discussed at chapter 4.2 the strong confinement scenario was taken into account in order to describe the Maxwell-Stefan diffusivities. Figure 8 presents the transient flux for the $CO_2$ at temperature 303 K. The feed pressures are 10, 100 and 1000 kPa respectively. From Figure 8 is extracted that as the feed pressure is increasing the $CO_2$ flux also reaches higher levels within the membrane. Further, the equilibrium time is low for all the three pressures ranging from 2 to 7 s and for the lower pressure of 10 kPa it seems that the equilibrium is reached faster than the higher pressures.

Figure 8. Transient flux of $CO_2$ at 303 K for three different feed pressures 10, 100 and 1000 kPa.

Figure 9 shows the transient profile for $CO_2$ permeation flux through the membrane across the z-axis of the zeolite membrane geometry which is the perpendicular axis to the membrane as shown in Figure 9d.

Figure 9. Transient $CO_2$ flux analysis across the z-axis of the zeolite geometry. Figure 9a shows the CO2 flux profile for feed pressure 10 kPa. Figure 9b shows the CO2 flux profile for feed pressure 100 kPa. Figure 9c shows the CO2 flux profile for feed pressure 1000 kPa and Figure 9d shows the z-axis of the zeolite membrane geometry which is the perpendicular axis to the membrane

The results showed that the flux profile has the same distribution for all the three feed pressures but the flux is higher as the feed pressure increases. The maximum flux value seems to be located in an area higher than the middle of the membrane. This distribution indicates that in the current geometry the flux of the $CO_2$ through the membrane has a standard profile for all the feed pressures and according the results of Figure 8 after some seconds the equilibrium is reached and the flux after the first 7s is constant. Figure 10 illustrates a three-dimensional representation of the $CO_2$ flux through the membrane where Figure 10a shows the results for feed pressure 10 kPa, Figure 10b shows the results for feed pressure 100 kPa and Figure 10c shows the results for feed pressure 1000 kPa respectively.

Figure 10. Three-dimensional representation of $CO_2$ flux with time and the z-axis of the zeolite membrabe. Figure 10a shows the results for feed pressure 10 kPa, Figure 10b shows the results for feed pressure 100 kPa and Figure 10c shows the results for feed pressure 1000 kPa respectively.

Figure 11 shows the transient flux of $CO_2$ at constant feed pressure 1000 kPa for three different temperatures 303, 363 and 423 K. According to the results extracted from Figure 11, the $CO_2$ flux through the membrane is higher for the temperature 303 K and as the temperature increases the flux seems to become lower. Further, it seems that for the lower temperature the equilibrium is reached slower than the case of the higher temperatures.

Figure 11. Transient flux of $CO_2$ at constant feed pressure 1000 kPa for three different temperatures 303, 363 and 423 K.

Figure 12 presents the transient analysis of the $CO_2$ flux across the z-axis of the zeolite membrane at constant feed pressure 1000 kPa and at three different temperatures 303, 363 and 423 K respectively and these results also indicates that for the current geometry there is a preferred distribution of the $CO_2$ flux through the membrane. This distribution is presented might due to the way that a Wicke-Kallenbach cell operates.

*Figure 12 Transient analysis of the $CO_2$ flux across the z-axis of the zeolite membrane at constant feed pressure 1000 kPa and at three different temperatures 303, 363 and 423 K. Figure 12a shows the flux for temperature 303 K. Figure 12b shows the flux for temperature 363 K and Figure 12c shows the flux for temperature 423 K.*

## 5   Conclusions

On the basis of the Maxwell-Stefan approach expressions have been derived for the description of the $CO_2$ diffusion through the zeolite membrane. A three dimensional study has been performed in a Wicke-Kallenbach cell with diameter of 19 mm and Argon used as the sweep gas in the permeate gas chamber. The proposed model was validated with experimental results and the similarity of the results was very good especially at higher temperatures maybe due to the fact that the support of the membrane might play a major role of thermal insulator at lower temperatures. Three different scenarios for the Maxwell-Stefan diffusion term were examined and compared to each other. The results showed that for high supply pressures only the strong and quasi-chemical approach have the potential in describing the $CO_2$ diffusion. The effect of temperature and pressure in the $CO_2$ permeation through the membrane was also studied and the results proved that for an almost linear isotherm the flux also can present linear behavior and the permeance is independent of the supply pressure. Finally, the transient analysis showed that the higher the supply pressure the higher the $CO_2$ flux through the membrane, but for lower pressures the time for reaching equilibrium state is lower. Further, for lower temperatures also the flux is greater comparing to higher temperatures but for the high temperatures the time for reaching equilibrium state is also lower.

# 6 Acknowledgements

The authors are very grateful to Professor Doros N. Theodorou (Head of COMSE Group, Department of Materials Science and Engineering, School of Chemical Engineering, National Technical University of Athens, Greece) for all fruitful discussions on this research work.

## Nomenclature

| | |
|---|---|
| $R$ | gas constant, *8.314 J/ mol/K* |
| $T$ | temperature, *K* |
| $u_i$ | velocity of species-i with respect to zeolite, *m/ s* |
| $D_{ij}$ | Maxwell-Stefan diffusivity describing interchange between *i* and *j*, *m² /s* |
| $D_i$ | Maxwell-Stefan diffusivity for species *i*, *m² s⁻¹* |
| $q_i$ | loading of component *i* in zeolite, *molecules per unit cell or mol kg⁻¹* |
| $N_i$ | Molecular flux of species-i, *molecules/m²/s or (mol/ m²/s)* |
| $p_i$ | Partial pressure of species-i, *Pa* |
| $B_{ij}$ | Elements of matrix [B], defined in Eq (10), *s /m²* |
| $c$ | Concentration of species, *mol/m³* |
| $d$ | density of $CO_2$, *kg /m³* |
| $n$ | dynamic viscosity, *Pa s* |
| $F$ | volume force, *force per unit volume* |
| $R$ | reaction rate, *1/s* |

*Greek Letters*

| | |
|---|---|
| $\mu$ | chemical potential, *J/ mol* |
| $\theta_i$ | fractional loading of component *i*, *dimensionless* |
| $\rho$ | density of membrane, *number of unit cells per m³ or kg /m³* |
| $\Gamma_{ij}$ | elements of the matrix of the thermodynamic correction factor [$\Gamma$], *dimensionless* |
| $\nabla$ | gradient operator |
| $\nabla^2$ | vector Laplacian |
| $\Pi$ | permeance, *mol/m²/s/Pa* |

*Superscripts*

| | |
|---|---|
| sat | referring to saturation loading |
| s | referring to surface diffusion |

**Figure Caption List**

Figure 1. Geometry of the Wicke-Kallenbach cell.

Figure 2. Comparison of experimental results extracted from recently published adsorption data by Himeno et al. and the simulation results extracted at same temperatures with a pressure step of 200 kPa (2a), and experimental adsorption data recently published by van der Bergh et al. and simulation results extracted at same temperatures with a pressure step of 10 kPa (2b).

Figure 3. Transient response to an increase in feed pressure in a $CO_2$ permeation system. Feed and Permeate fluxes for the three different scenarios about the M-S diffusivities for feed pressures a) 10 kPa b) 100 kPa and c) 1000 kPa.

Figure 4. $CO_2$ flux as function of the feed pressure at constant temperature 303 K.

Figure 5. Adsorption isotherms of carbon dioxide at 195, 252, 273 and 298 K.

Figure 6. Effect of pressure and temperature on the permeation of carbon dioxide through the zeolite membrane.

Figure 7. Comparison of the results extracted from simulation runs and the experimental results extracted from van der Broeke et al. [8] for the effect of pressure and temperature on the permeance

Figure 8. Transient flux of $CO_2$ at 303 K for three different feed pressures 10, 100 and 1000 kPa.

Figure 9. Transient $CO_2$ flux analysis across the z-axis of the zeolite geometry. Figure 9a shows the CO2 flux profile for feed pressure 10 kPa. Figure 9b shows the CO2 flux profile for feed pressure 100 kPa. Figure 9c shows the CO2 flux profile for feed pressure 1000 kPa and Figure 9d shows the z-axis of the zeolite membrane geometry which is the perpendicular axis to the membrane

Figure 10. Three-dimensional representation of $CO_2$ flux with time and the z-axis of the zeolite membrabe. Figure 10a shows the results for feed pressure 10 kPa, Figure 10b shows the results for feed pressure 100 kPa and Figure 10c shows the results for feed pressure 1000 kPa respectively.

Figure 11. Transient flux of $CO_2$ at constant feed pressure 1000 kPa for three different temperatures 303, 363 and 423 K.

Figure 12. Transient analysis of the $CO_2$ flux across the z-axis of the zeolite membrane at constant feed pressure 1000 kPa and at three different temperatures 303, 363 and 423 K. Figure 12a shows the flux for temperature 303 K. Figure 12b shows the flux for temperature 363 K and Figure 12c shows the flux for temperature 423 K.

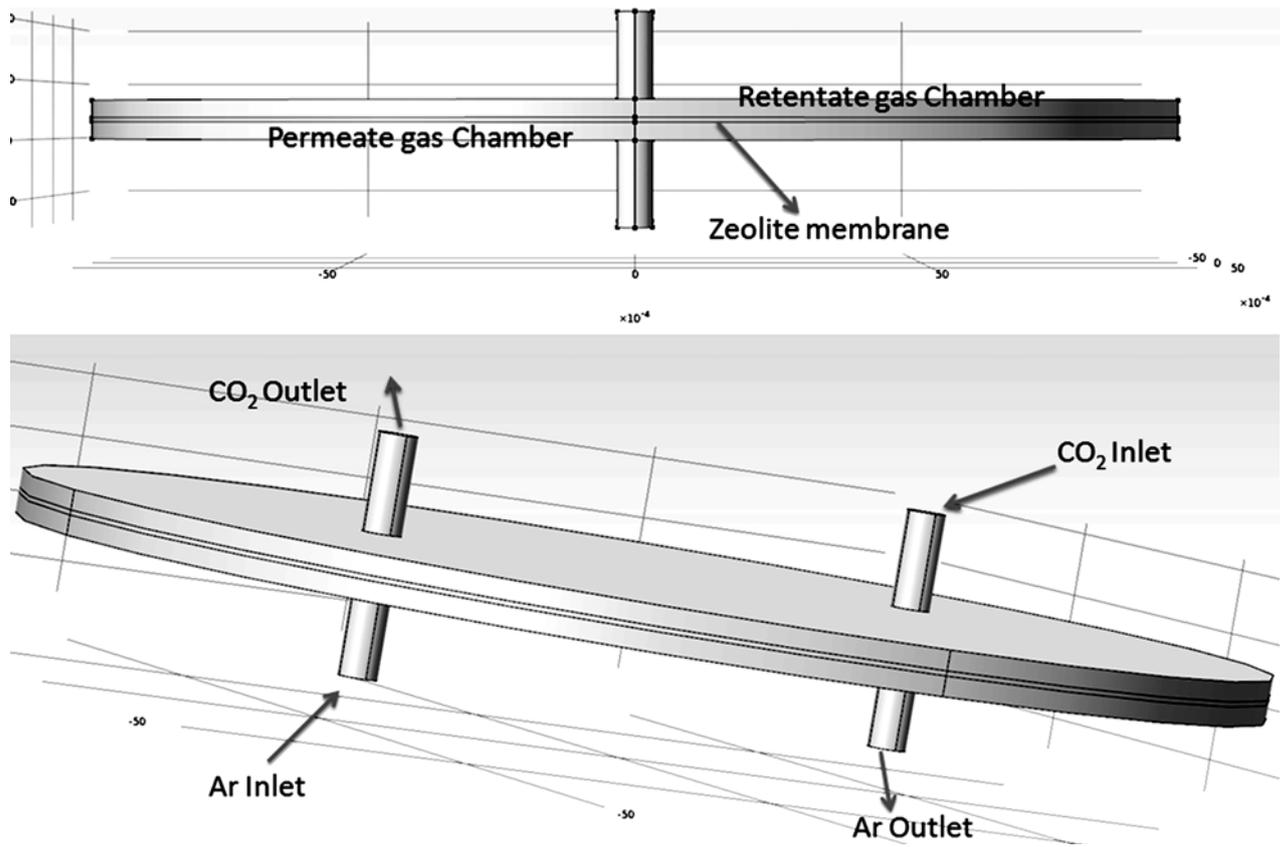

Figure 1. Geometry of the Wicke-Kallenbach cell

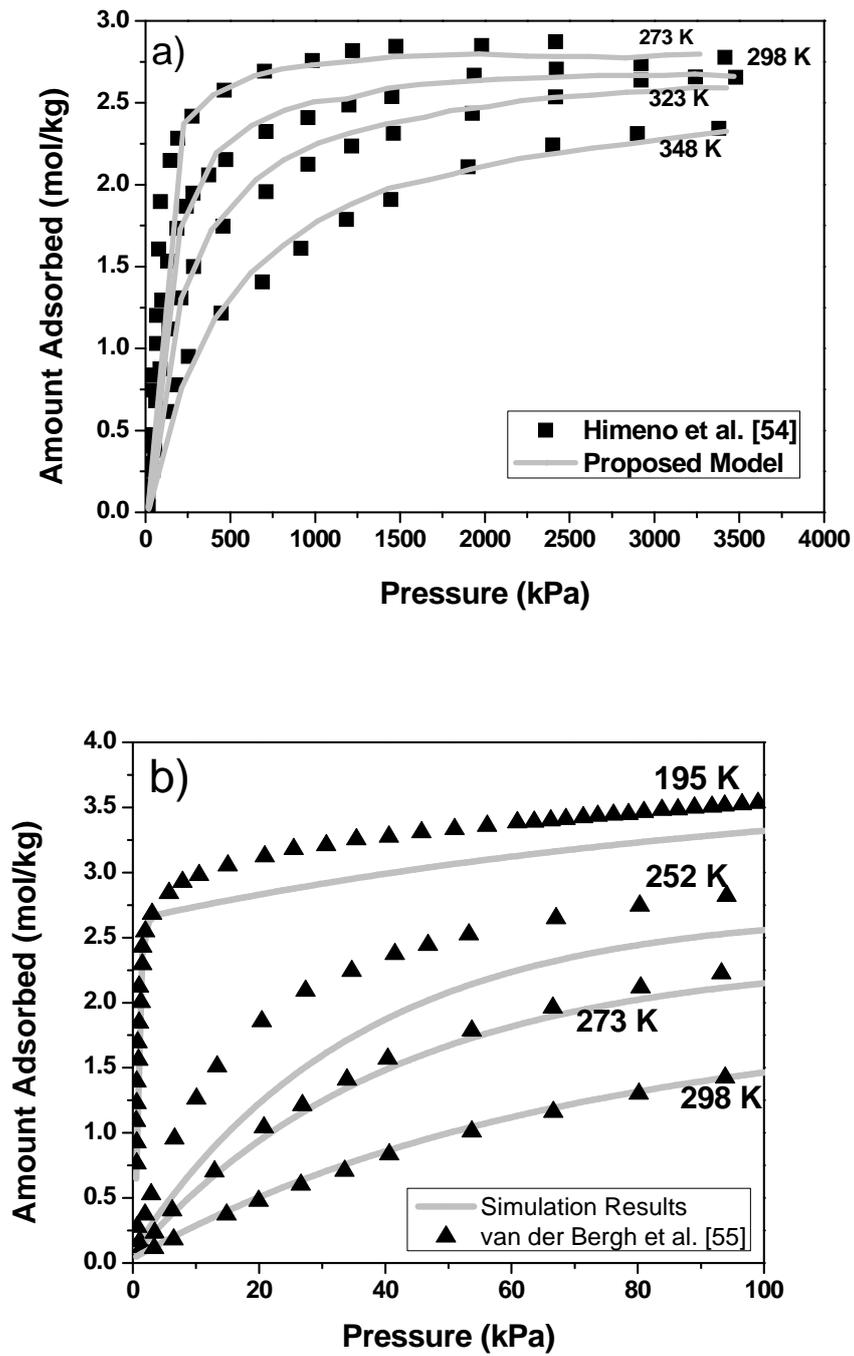

Figure 2. Comparison of experimental results extracted from recently published adsorption data by Himeno et al. and the simulation results extracted at same temperatures with a pressure step of 200 kPa (2a), and experimental adsorption data recently published by van der Bergh et al. and simulation results extracted at same temperatures with a pressure step of 10 kPa (2b).

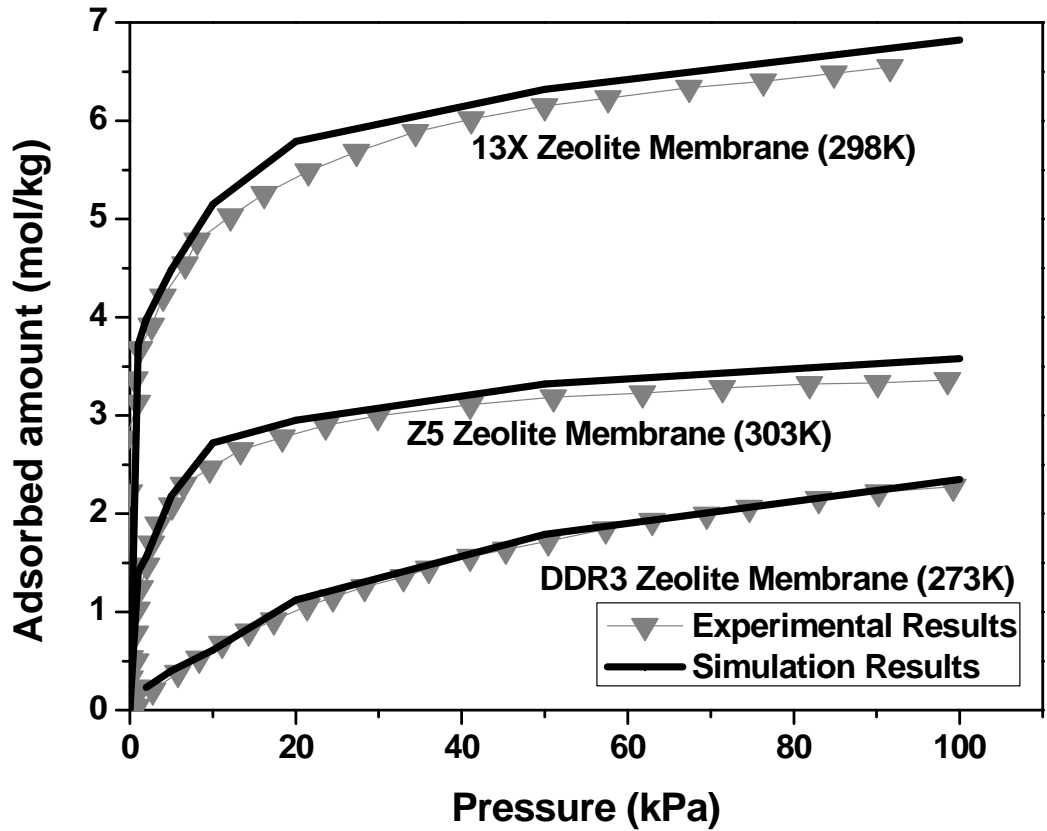

Figure 3. Comparison of experimental results extracted from recently published adsorption data for three different type of zeolite membranes, DDR3 membrane by van den Bergh et al. [55], Z5 membrane by Liu et al. [58] and 13X membrane by Cavenati et al. [59] and the simulation results extracted at 273 K for the DDR3 membrane, 303 K for the Z5 membrane and 298 K for the 13X membrane

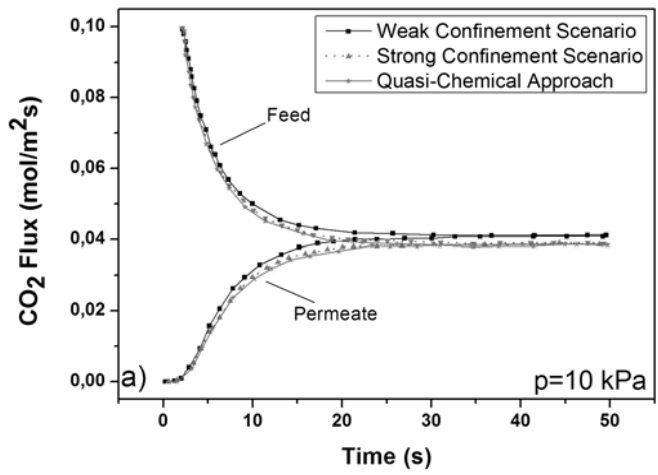
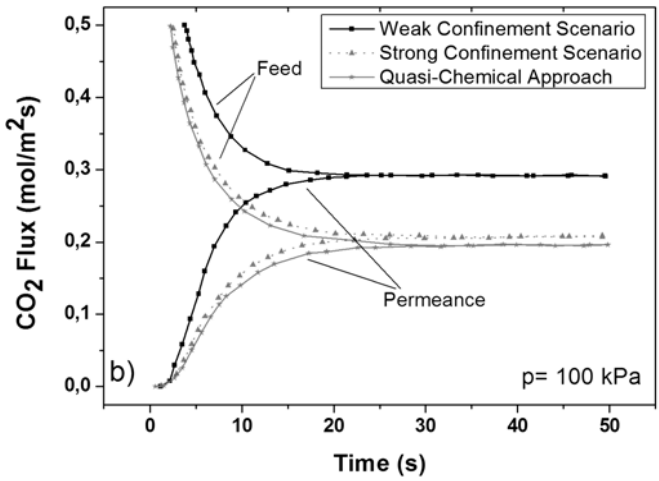
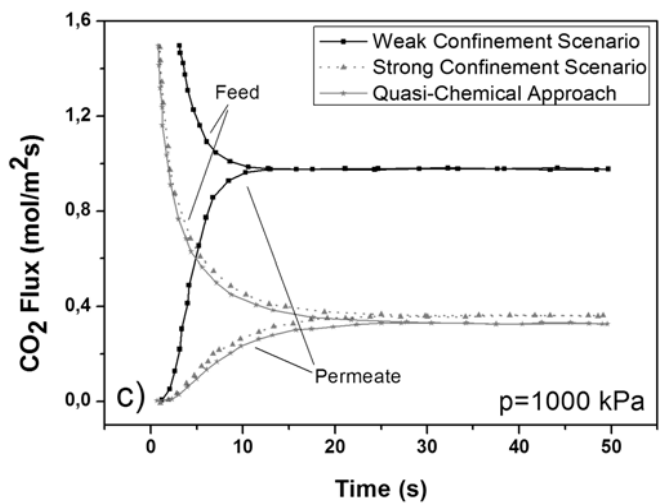

Figure 4. Transient response to an increase in feed pressure in a $CO_2$ permeation system. Feed and Permeate fluxes for the three different scenarios about the M-S diffusivities for feed pressures a) 10 kPa b) 100kPa and c) 1000 kPa.

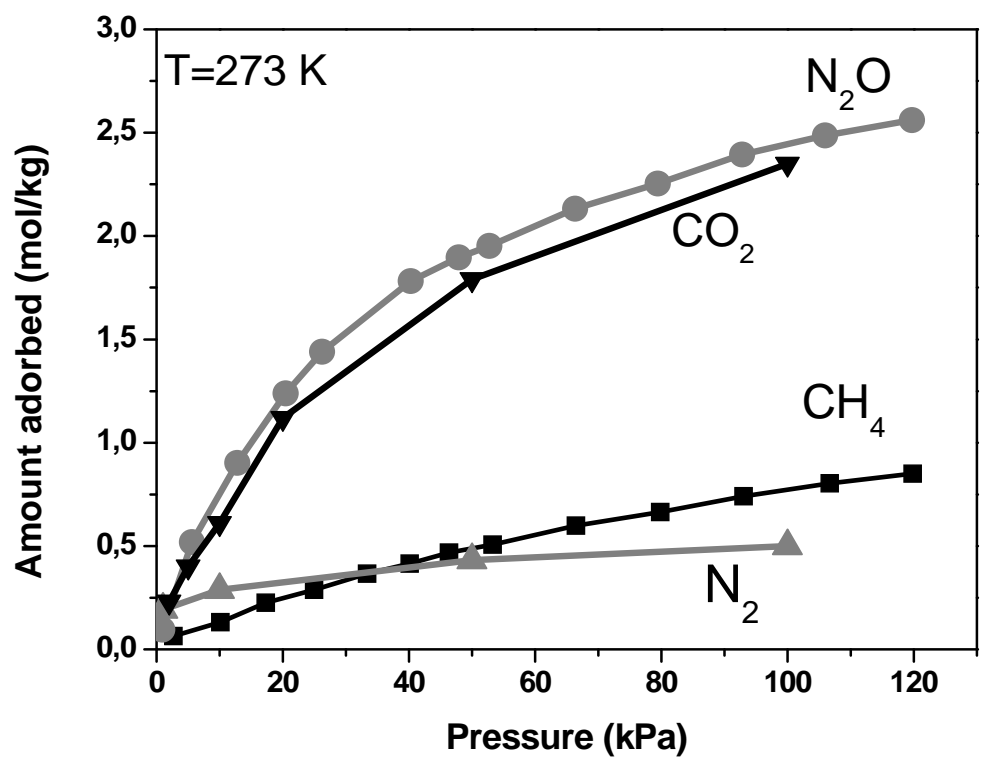

Figure 5 Adsorption isotherms for carbon dioxide, nitrous oxide, methane and nitrogen in DDR3 zeolite membrane at 273 K.

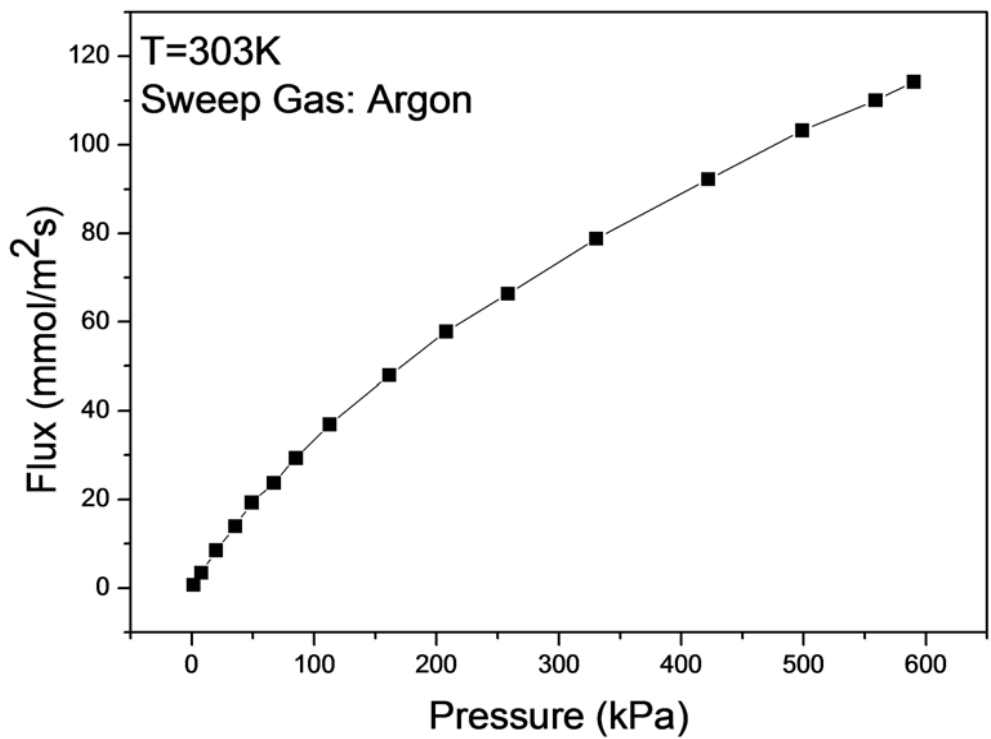

Figure 6. $CO_2$ flux as function of the feed pressure at constant temperature 303 K.

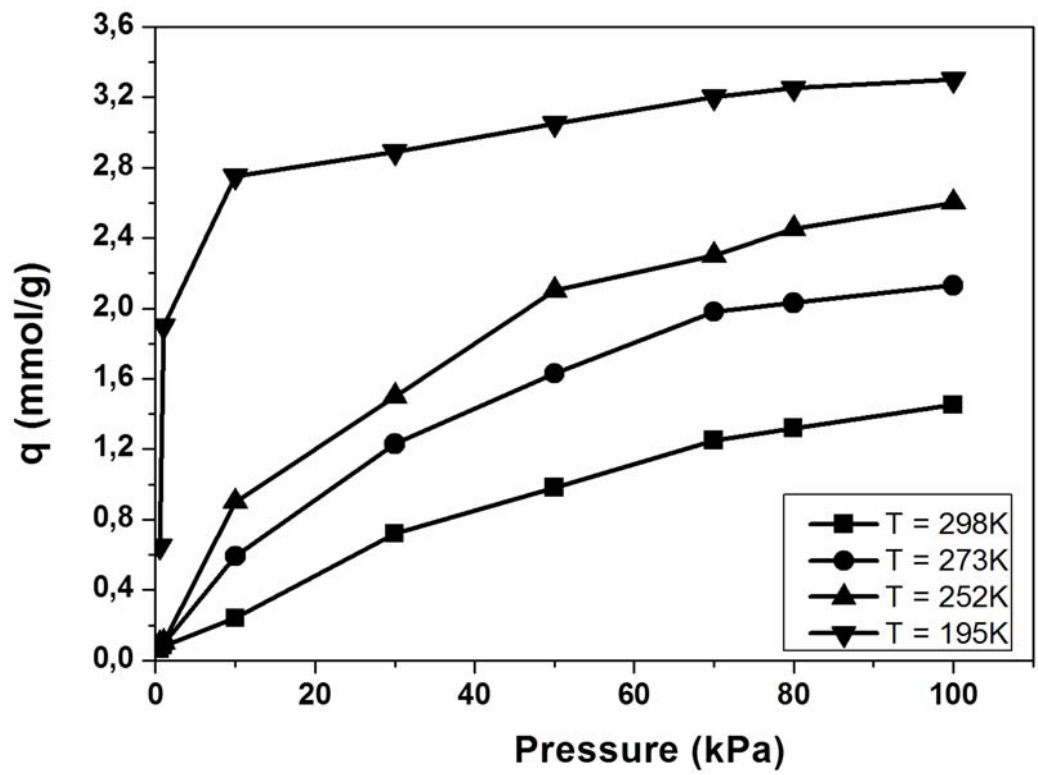

Figure 7. Adsorption isotherms of carbon dioxide at 195, 252, 273 and 298 K.

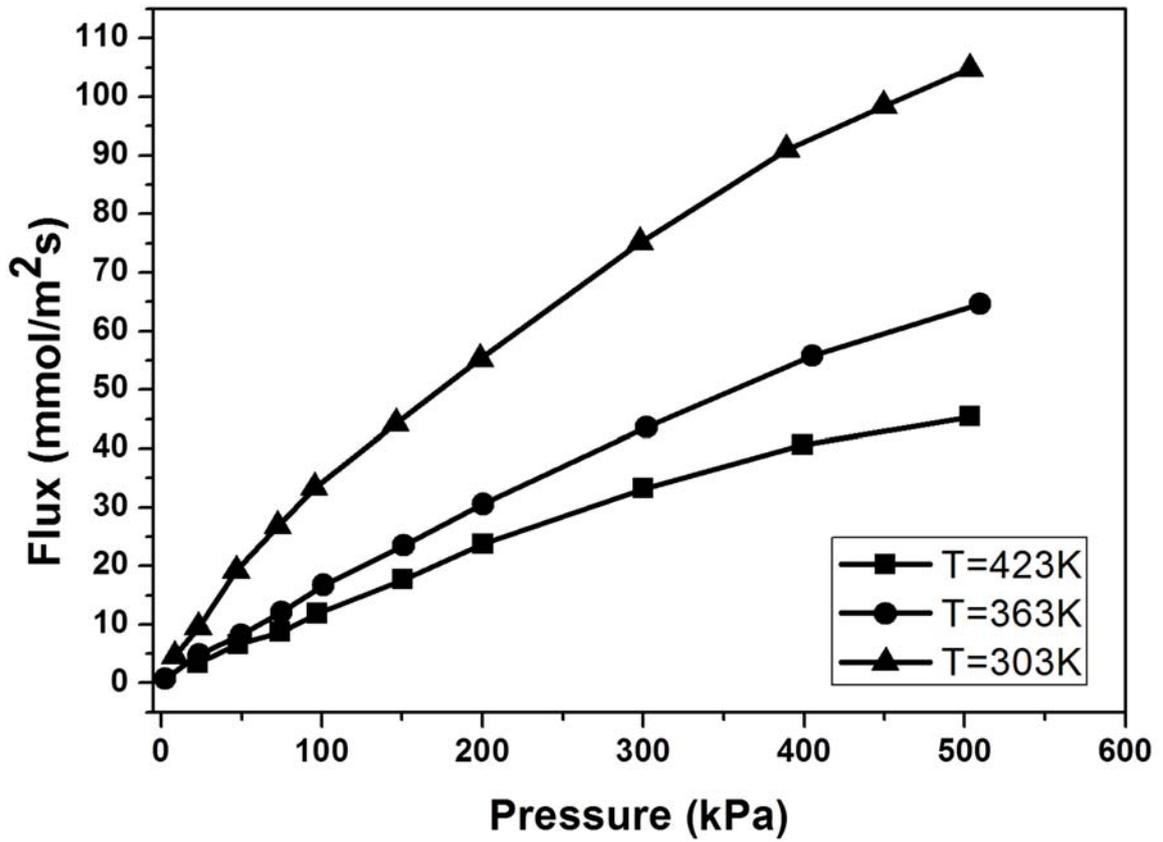

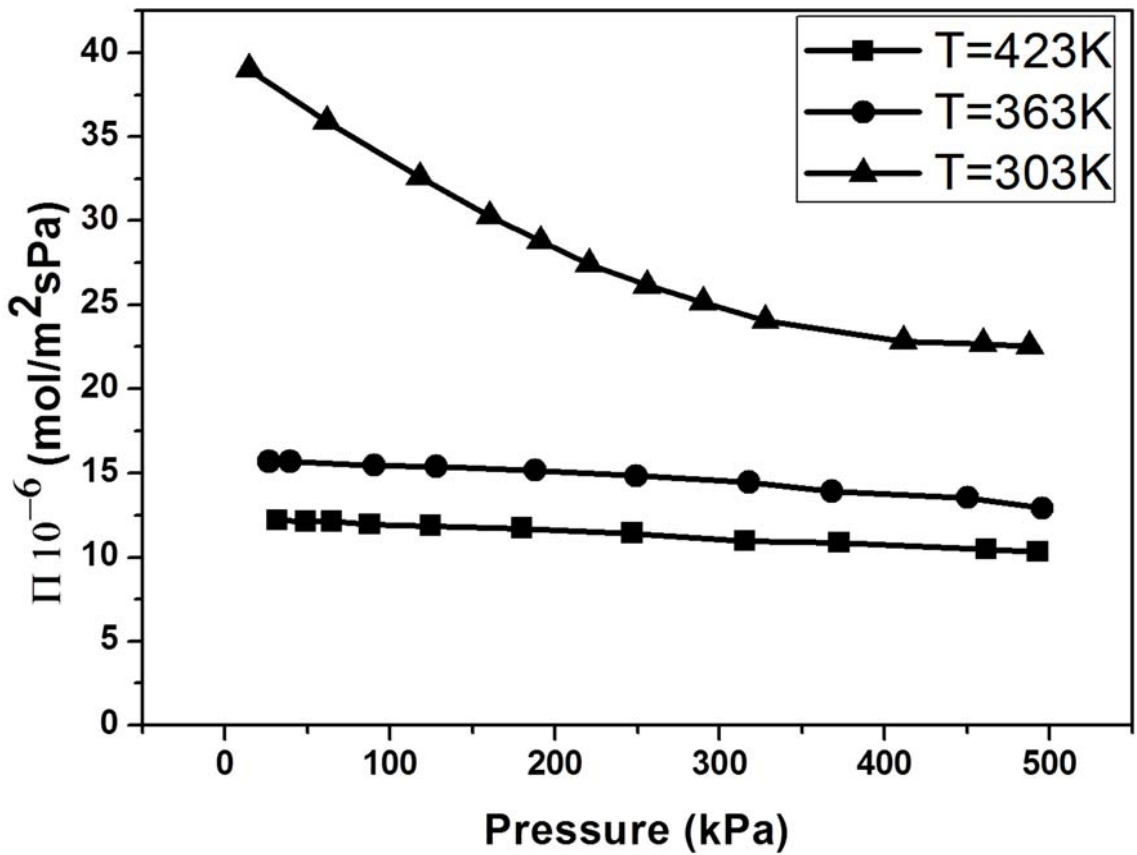

Figure 8. Effect of pressure and temperature on the permeation of carbon dioxide through the zeolite membrane.

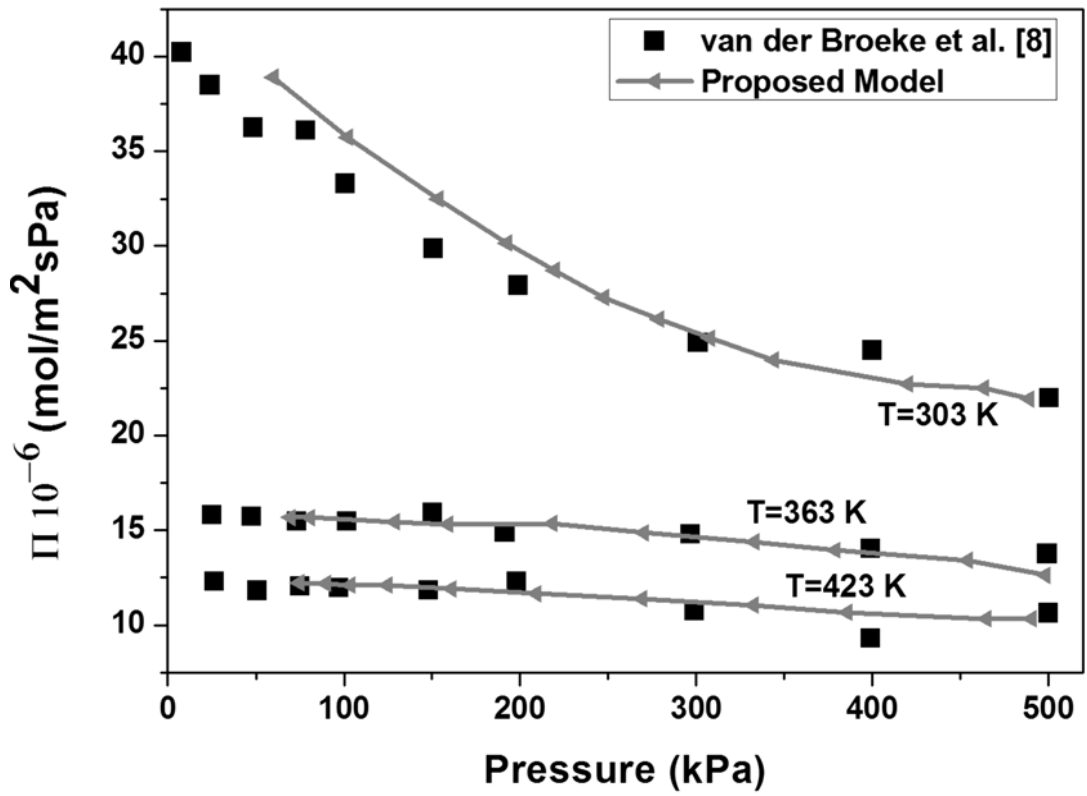

Figure 9. Comparison of the results extracted from simulation runs and the experimental results extracted from van der Broeke et al. [8] for the effect of pressure and temperature on the permeance

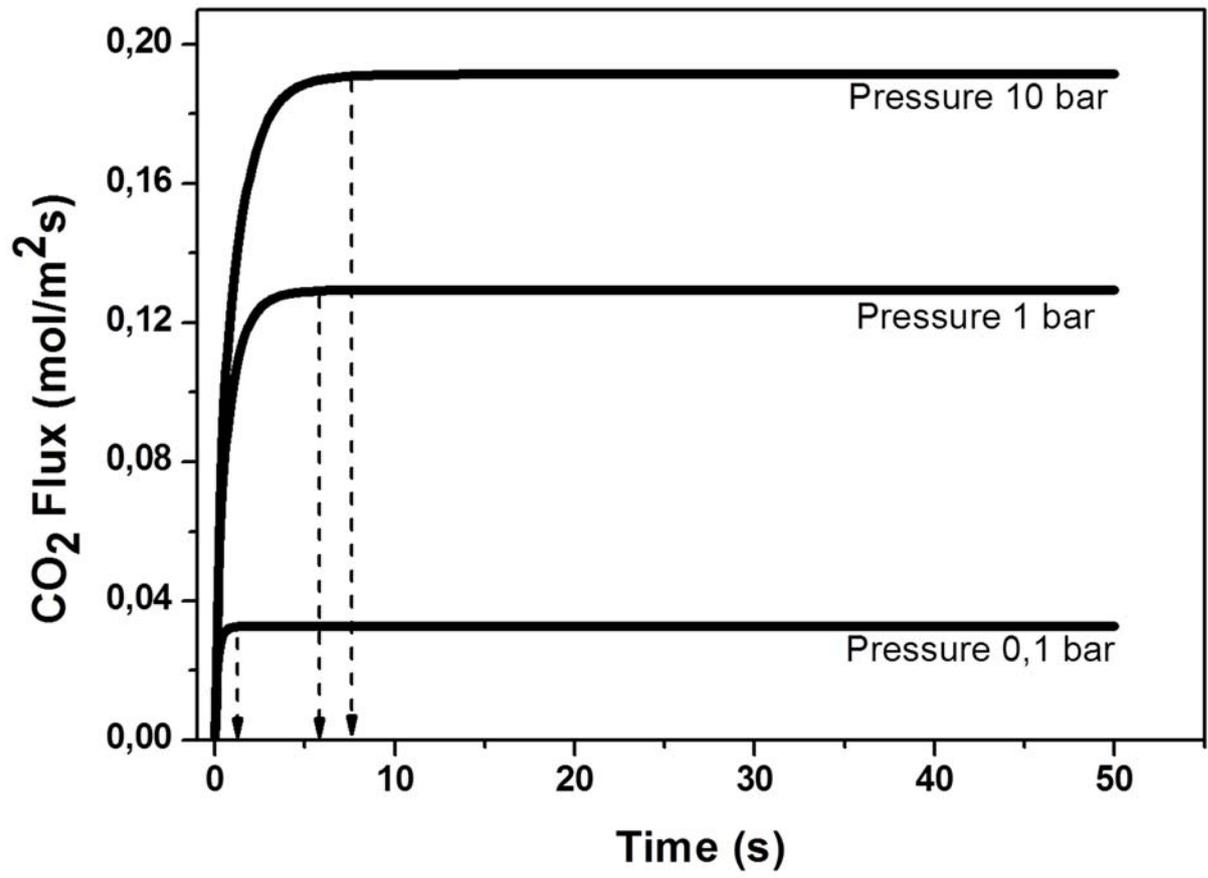

Figure 10. Transient flux of $CO_2$ at 303 K for three different feed pressures 10, 100 and 1000 kPa.

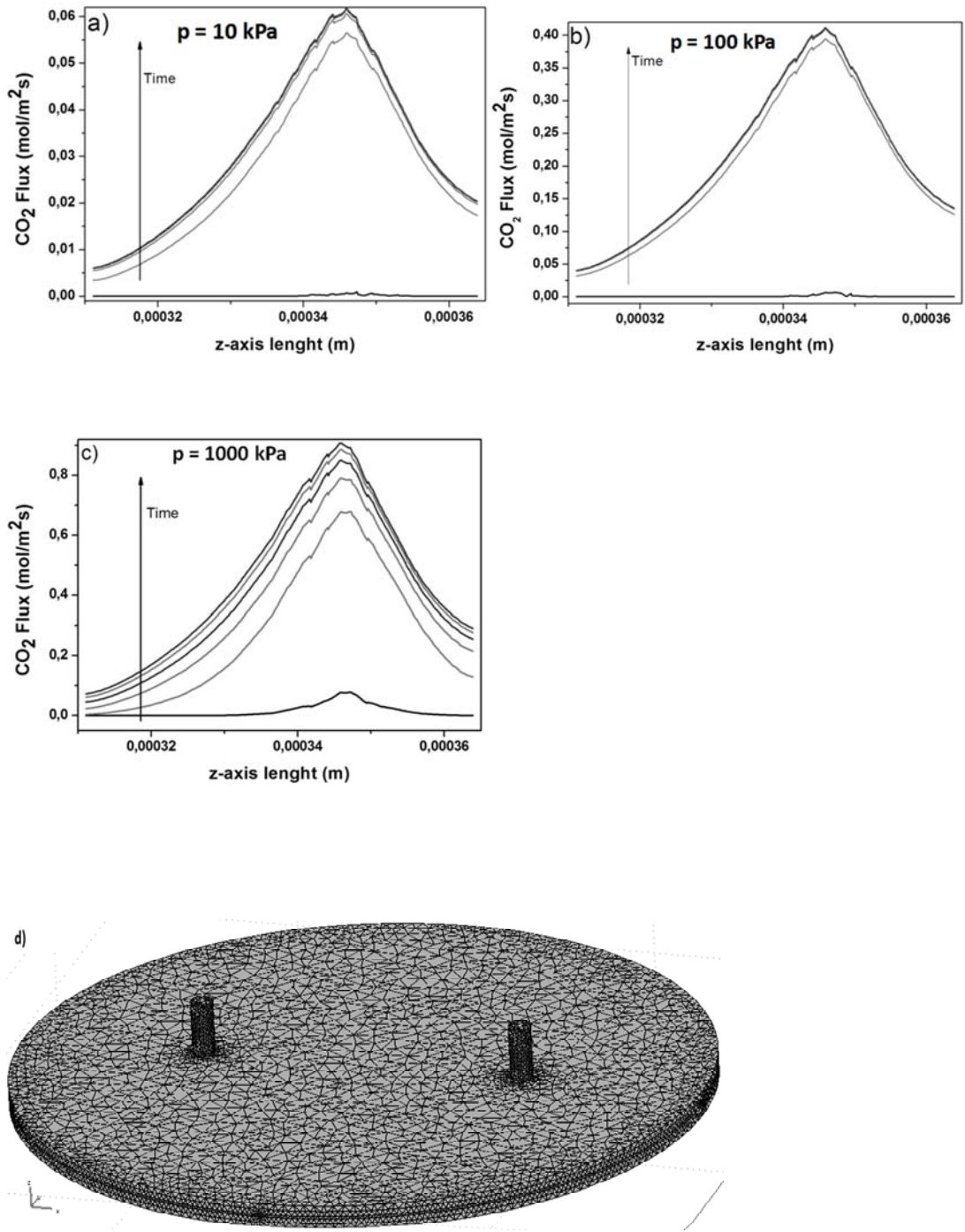

Figure 11. Transient CO$_2$ flux analysis across the z-axis of the zeolite geometry. Figure 9a shows the CO2 flux profile for feed pressure 10 kPa. Figure 9b shows the CO2 flux profile for feed pressure 100 kPa. Figure 9c shows the CO2 flux profile for feed pressure 1000 kPa and Figure 9d shows the z-axis of the zeolite membrane geometry which is the perpendicular axis to the membrane

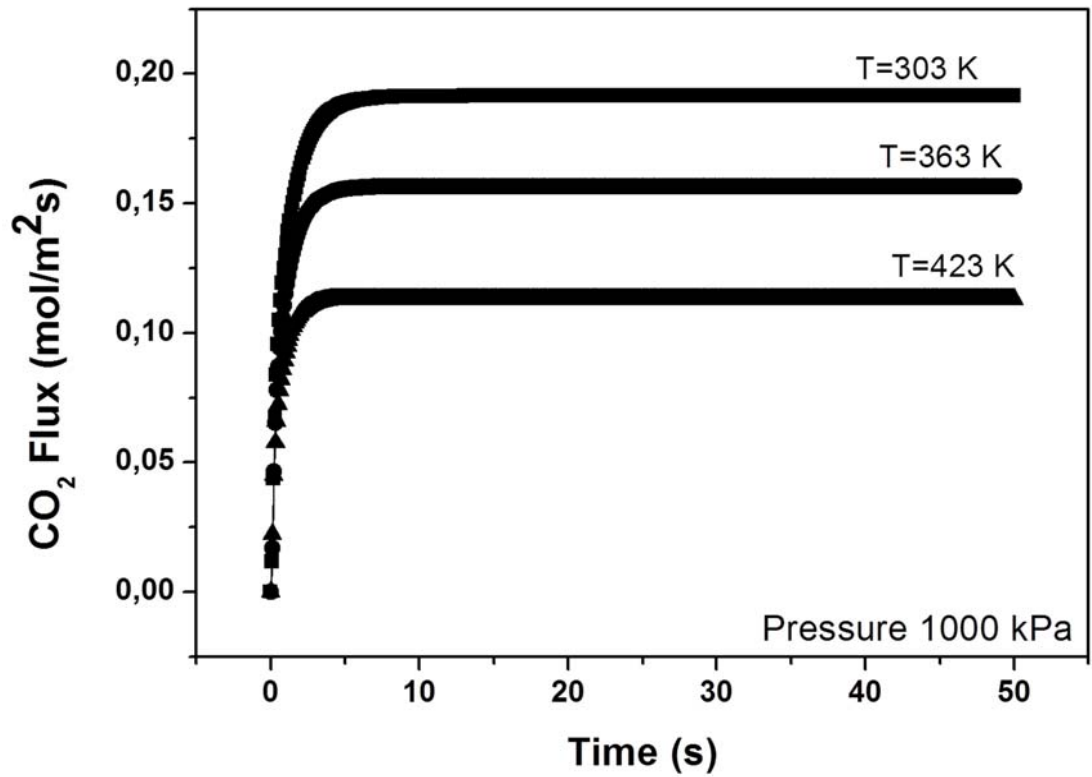

Figure 12. Transient flux of $CO_2$ at constant feed pressure 1000 kPa for three different temperatures 303, 363 and 423 K.

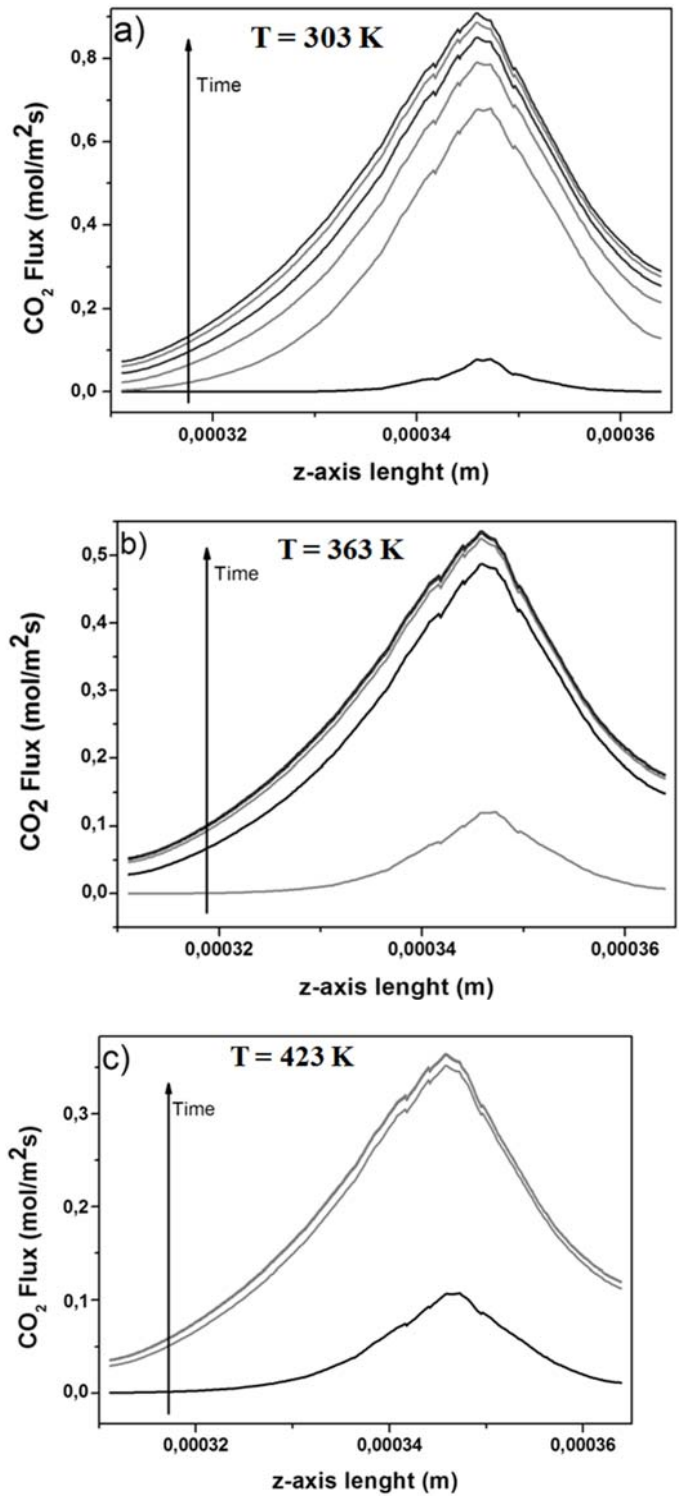

Figure 13. Transient analysis of the $CO_2$ flux across the z-axis of the zeolite membrane at constant feed pressure 1000 kPa and at three different temperatures 303, 363 and 423 K. Figure 12a shows the flux for temperature 303 K. Figure 12b shows the flux for temperature 363 K and Figure 12c shows the flux for temperature 423 K.

Table 1. Parameters used for the comparison of the three scenarios [37]

| Species | $q_i^{sat}$ (mol/kg) | $\Delta H_{ad}$ | $D_i(0)\,(m^2/s)$ |
|---|---|---|---|
| $CO_2$ | 4.161 | -25000 | $5.277 \times 10^{-9}$ |

Table 2. Langmuir parameters for the temperatures 303, 363 and 423 K.

| T(K) | $q^{sat}$ (mol/kg) |
|---|---|
| 303 | 2.48 |
| 363 | 1.81 |
| 423 | 1.46 |